\documentclass{article}

\usepackage{amsmath}
\usepackage{amssymb}
\usepackage[margin=1in]{geometry}
\usepackage{graphicx}
\usepackage{caption}
\usepackage{lscape}
\usepackage{setspace}
\usepackage{changepage}
\usepackage[affil-it]{authblk}
\usepackage{apacite}
\usepackage[T1]{fontenc}
\usepackage{newtxmath,newtxtext}
\usepackage{hyperref}
\usepackage{breakurl}

\usepackage[round]{natbib} 
\DeclareMathOperator*{\argmax}{argmax}

\makeatletter

\begin{document}
	\title{MACROECONOMICS AND FINTECH: UNCOVERING LATENT MACROECONOMIC EFFECTS ON PEER-TO-PEER LENDING}
	\author[1]{JESSICA FOO}
	\author[2]{LEK-HENG LIM}
	\author[1]{KEN SZE-WAI WONG\thanks{Corresponding to: Ken Sze-Wai Wong, Department of Statistics, University of Chicago, 5747 South Ellis Avenue Chicago, IL 60637, USA.  Email: \url{kenwong@uchicago.edu}}}
	\affil[1]{Department of Statistics, University of Chicago}
	\affil[2]{Computational and Applied Mathematics Initiative, Department of Statistics, University of Chicago}

	\maketitle

\doublespacing
	
	\begin{abstract}
		Peer-to-peer (P2P) lending is a fast growing financial technology (FinTech) trend that is displacing traditional retail banking. Studies on P2P lending have focused on predicting individual interest rates or default probabilities. However, the relationship between aggregated P2P interest rates and the general economy will be of interest to investors and borrowers as the P2P credit market matures. We show that the variation in P2P interest rates across grade types are determined by three macroeconomic latent factors formed by Canonical Correlation Analysis (CCA) --- macro default, investor uncertainty, and the fundamental value of the market. However, the variation in P2P interest rates across term types cannot be explained by the general economy.

	\end{abstract}

\section{Introduction} \label{sec:intro}
Online marketplace lending, also known as peer-to-peer (P2P) lending, directly connects borrowers with lenders on an online platform, bypassing traditional financial intermediaries such as banks. Borrowers apply for loans, specifying loan details which are then listed on the platform. Prospective lenders can then access the platform and view listings, choosing to fund the loan fully or partially. The interest rate that corresponds to the individual loan is determined by the platform, which collects personal data on the individual to determine credit worthiness, before assigning a loan grade to the borrower. Loans with a higher credit risk will thus entail a higher interest rate, due to the higher possibility of late repayment or default. While traditional banks rely largely on FICO credit scores to determine loan amounts and rates, P2P platforms use complex algorithms to analyze personal data ranging from social media activity to platform use to determine credit worthiness. Furthermore, these algorithms adjust interest rates based on the demand and supply of P2P loans, functioning like a price mechanism.

P2P lending is part of the rapidly growing financial technology (FinTech) industry that is transforming and disrupting traditional financial services. Besides eliminating inefficiencies and overhead costs of transactions between borrowers and lenders, P2P lending allows borrowers with low FICO scores to obtain loans that were previously inaccessible to them. P2P lending is also advantageous for lenders, offering the possibility of higher yields for investors who wish to diversify their portfolio. Given these advantages, Transparency Market Research anticipates the global P2P market to expand at a compounded annual growth rate (CAGR) of 48.2\% between 2016 and 2024. In the U.S., P2P lending has grown an average of 84\% per quarter from  2007 to 2014.\footnote{Peer-to-Peer Lending is Poised to Grow. \textit{Federal Reserve Bank of Cleveland}. August 14, 2014.} However, while banks, credit card companies and other traditional lending companies generate more than \$870 billion annually from fees and interest, P2P's 2014 lending revenue constitutes only about 2\% of that amount,\footnote{A Trillion Dollar Market By the People, For the People. \textit{Foundation Capital}. May 6, 2014.} with Prosper and Lending Club capturing 98\% of the market.\footnote{Banking Without Banks. \textit{The Economist}. February 28, 2014.} Nevertheless, P2P lending is projected to service 10\% to 15\% of consumer debt by 2025.\footnote{Peer Pressure: How Peer-to-Peer Lending Platforms are Transforming the Consumer Lending Industry. \textit{PwC Consumer Finance Publications}. February, 2015.}

Given the potential of P2P lending to disrupt retail banking, this paper aims to analyze the relationship between P2P lending and the general economy. Economic theory suggests that credit markets are correlated with macroeconomic activity as they are used to channel an economy's savings into other more productive uses. Analyzing the interest rates of different loans can thus reveal investors' expectations of the economy, which are correlated with economic variables like unemployment rate and inflation. The corporate credit spread, defined as the additional yield investors receive for investing in a corporate bond instead of a government bond of similar term (loan period), has been found to be correlated with the economy. In this paper, we attempt to analyze the P2P credit spread, which we find analagous to the corporate credit spread, since both involve differing maturity terms and loan grade types, and adopt the methodologies used by \citet{ahn} in analyzing corporate credit spreads.

Since macroeconomic variables are proxies of the economy, we construct latent factors from canonical correlation analysis (CCA) of P2P credit spreads and observed explanatory macroeconomic variables, and investigate the extent to which the factors capture the systematic variation of P2P credit spreads. We base our initial selection of macroeconomic variables on economic theory and intuition, and rely on them to interpret the factors.

We show that the common variation in P2P credit spreads is mainly explained by three factors. The first factor is a macro non-default factor, which is strongly correlated with economic indicators of economic expansion such as increased inflation, reduced unemployment and reduced household debt. Furthermore, an increase in the first factor is associated with an increase in P2P credit spreads of all term and grade types in a CCA factor regression. We identify and interpret the second factor as a latent market uncertainty factor, which is positively correlated with the risk-free slope and equity volatility. An increase in the second factor is associated with a decrease in P2P credit spreads across grade types, and not significantly associated in P2P credit spreads across term types. Our third factor can be interpreted as the fundamental value of the market, although canonical cross-loadings for the third factor are weak.

We also find that unlike corporate bonds or government bonds, P2P loans have little term structure that can be explained by macroeconomic conditions. One limitation is that P2P loans are currently only offered on two terms --- 36 months and 60 months, making it difficult to observe interest rate variation across term types and the associated risk with longer term loans. However, across grade types, we find that macroeconomic factors do adequately explain P2P credit spread variation.

Our paper proceeds as follows: Section~\ref{sec:litreview} describes existing literature related to P2P lending and credit markets. Section~\ref{sec:method} outlines our methodology. Section~\ref{sec:data} describes our data set and selection of macroeconomic proxy variables. Section~\ref{sec:results} shows our results using multiple regressions, canonical correlation factors and our economic interpretations. We conclude in Section~\ref{sec:conclusion}, and the Appendix contains Tables and Figures. 

\section{Literature Review} \label{sec:litreview}
Academic research on P2P lending has focused primarily on two aspects --- the determinants of funding success, and of default. The predictors studied are detailed borrower-specific variables that include personal information, credit history, financial ability and social information. To predict funding success or default, studies have focused on classification models, which either classify loans into loan sub-grades ranging from A to H (with A being the most credit worthy) or into binary predictions of default.

\citet{klafft} found that funding success on Prosper is significantly influenced by verified bank account information and credit rating, while interest rates are primarily influenced by credit rating and debt-to-income ratios. Similarly, using a multivariate logistic regression model, \citet{herzenstein} found that funding success on Prosper depends more on borrowers' financial strength, such as credit grade and debt-to-income ratio, and their efforts to obtain the loan (providing detailed information or joining platform communities) than demographic features such as race. \citet{zhang} used a decision tree to conclude that loan information, social media information and credit information are most crucial in determining default risk on China's leading P2P platform, PPDai. In addition to loan-specific information and borrower-specific information, \citet{dietrich} included macroeconomic variables into ordinary least squares regression and found that higher unemployment rates and three-year government bond yield rates were correlated with higher P2P lending rates in Switzerland. In essence, most research related to P2P lending have used individual data to predict individual loan interest rates.

On the other hand, traditional credit markets have been frequently analyzed using aggregate loan data, and have also relied primarily on factor models. \citet{litterman} first used principal component analysis on the yields of U.S.\ government bonds, and found that most of the variations in returns can be explained by first three eigenvectors, also known as yield curve factors in empirical finance, which they called level, steepness and curvature. Yield changes caused by the level factor are constant across maturities, resulting in a parallel shift in the yield curve, while the steepness factor raises the yields of treasuries with longer maturities, compared to those with shorter maturities. However, while no meaningful interpretation was provided for the third factor, they found that changes in the third factor is correlated with changes in interest rate volatility. \citet{ang} then analyzed the latent factors and macroeconomic variables jointly in a vector autoregression, and found that forecasting performance was better for models that included additional macro factors, than models with only latent variables. Exploring the relation between latent factors and macro variables, \citet{ahn} used canonical correlations between corporate credit spreads and macroeconomic variables to form factors that explain variation in the term structure of corporate credit spreads. They found three common factors that accounted for 40\% of the total variation, with the first related to the contemporaneous state of the economy, the second to expectations of future economic conditions, and the third to error correction in short-term spreads.

\section{Methodology} \label{sec:method}

\subsection{Basic Model} \label{sec:basicmodel}
We explore the theoretical model in \citet{ahn} that takes into account two possible scenarios --- first, that there are missing non-macroeconomic related factors not captured by macroeconomic indicators, and second, that the macroeconomic indicators may not be perfect proxies for true macroeconomic factors, resulting in ``systematic measurement errors'' that can be found in the residuals. To demonstrate this, we first assume that each $n$-dimensional credit spread variable $y_t$ follows a linear $m$-factor model given by 
\begin{equation} \label{factor}
y_t = \alpha + B_1 f_{1,t} + B_2 f_{2,t} + u_t
\end{equation}
where $\alpha$ is a $n \times 1$ vector of intercepts, $u_t$ is a $n \times 1$ vector of zero-mean idiosyncratic components, $B_1$ and $B_2$ are $n \times r$ and $n \times (m-r)$ factor loading matrices, $f_{1,t}$ and $f_{2,t}$ are $r \times 1$ and $(m-r) \times 1$ vectors of common factors. We assume that systematic variation in the dependent variables can all be captured by $m$ common factors. The factors $f_{1,t}$ are predicted by a theoretical model of interest, which in our case relates the P2P credit spreads with the wider economy. The factors $f_{2,t}$, however, are not predicted by the macro-economy. Both factors $f_{1,t}$ and $f_{2,t}$ are unobserved, but we observe $k$ proxy macroeconomic variables that are correlated with factors $f_{1,t}$, and will refer to these proxy variables as $z_t$. We assume further that the factors in $f_{2,t}$ are uncorrelated with the proxy variables $z_t$. Our formal assumptions are thus: 
\begin{enumerate}
\item $\mathbb{E} [u_t, (f_{1,t}', f_{2,t}', z_t') ] = 0_{n \times m + k}.$ Hence all common factors and proxy variables are uncorrelated with idiosyncratic error terms in $u_t$. 
\item $\operatorname{rank}(B_1) = r$; $\operatorname{rank}(B_2) = m-r.$ Hence the model is fully specified with $m$ factors.
\item The linear projections of $f_{1,t}$ and $f_{2,t}$ on $z_t$ are
\begin{equation}
\begin{split}
L(f_{1,t} \mid z_t) &= \Theta' z_t \\ L(f_{2,t} \mid z_t) &= 0_{m-r \times 1}
\end{split}
\end{equation}
Hence only the factors $f_{1,t}$ are correlated with macroeconomic proxies
\item $\operatorname{Var}(u_t) \equiv \Omega_{uu}$ is a diagonal matrix. Therefore, conditional on the factors, there is no serial correlation in $y_t$
\end{enumerate}
Therefore, substituting the linear projections in equation (2) into equation (1) with some algebraic manipulation, we get 
\begin{equation} \label{linear}
\begin{split}
y_t &= \alpha + B_1 [f_{1,t} + L(f_{1,t}) - L(f_{1,t})] + B_2 f_{2,t} + u_t \\ &= \alpha + B_1  \Theta' z_t + B_1 [f_{1,t} - L(f_{1,t})] +  B_2 f_{2,t} + u_t \\ &= \alpha + B_1  \Theta' z_t + B_1 v_t +  B_2 f_{2,t} + u_t \\ &=  \alpha + B_1  \Theta' z_t + e_t
\end{split}
\end{equation} 
\newline
From  \eqref{linear}, it is evident that the residuals, $e_t$ can contain two sources of systematic variation. The first is due to a missing factors $f_{2,t}$ that are not correlated with the proxies $z_t$. The second is due to incomplete specification of $z_t$, resulting in imperfection correlation between $z_t$ and $f_{1,t}$, which implies that $v_t$, the projection errors, are not 0. Therefore, if we regress $y_t$ on $z_t$ with ordinary least squares, the residuals will contain factor structure that can be uncovered with principal components, in both scenarios. However, if we adopt CCA factors as in \citet{ahn}, we can determine whether there are missing factors unaccounted by macroeconomic variables. 

\subsection{CCA Factor Model} \label{sec:ccafactor}
Canonical correlation analysis (CCA) allows us to analyze linear relationships between two sets of random variables, by finding the optimal canonical variates --- the linear combinations of the variables within each set, which maximize the correlation between them.

Given $p$ variables in one set, and $q$ variables in another set, let $Y = (Y_1', Y_2', \dots, Y_p')$ be the first set and let $Z = (Z_1', Z_2', \dots, Z_q')$ be the second set. Then, consider the linear combinations $U_i = a_i'Y \ \text{for all } i \in \{1,\dots,p\}$ and $V_j = b_j'Z$ for all $j \in \{1,\dots,q \}$, where $a_i \in \mathbb{R}^p$ and $b_j \in \mathbb{R}^q$. Define
\begin{equation}
\operatorname{Corr}(U_i, V_j) = \frac{a_i' \Sigma_{YZ} b_j}{\sqrt{a_i' \Sigma_Y a_i} \sqrt{b_j' \Sigma_Z b_j}}
\end{equation}
where $\Sigma_Y = \operatorname{Var}(Y), \Sigma_Z = \operatorname{Var}(Z), \Sigma_{YZ} = \operatorname{Cov}(Y, Z)$. Then the first pair of canonical variates are $(U_1, V_1) = \argmax \{\operatorname{Corr}(U_1, V_1): \operatorname{Var}(U_1) = \operatorname{Var}(V_1) = 1 \}$. In general, the $k$th pair of canonical variate has additional constraints that the $k$th canonical variates have to be uncorrelated with the other $1,\dots,k-1$ canonical variates. Therefore, the $k$th pair of canonical variates is defined as:
\begin{multline*}
(U_k, V_k) = \argmax \{\operatorname{Corr}(U_k, V_k):  \operatorname{Var}(U_k) = \operatorname{Var}(V)_k = 1, \\ \operatorname{Cov}(U_k, U_j) = \operatorname{Cov}(V_k, V_j) = \operatorname{Cov}(U_k, V_j),\\ \operatorname{Cov}(V_k, U_j) = 0\; \text{for all} \; j = 1,\dots,k-1 \}.
\end{multline*}
Assuming that $Y$ represents the set of P2P yields, and $Z$ the set of macroeconomic variables, we then use $(U_1, U_2, \dots, U_r)'$ as CCA factors, with $r$ being the number of factors chosen.

\citet{ahn} have shown that using $U = A'Y$ as estimated factors, where $A = (a_1,\dots,a_r)$, has two main advantages. Firstly, the factor loadings based on the CCA factors are the same as the factor loadings of the true factors $f_{1,t}$ up to a linear transformation. This is because the columns of $\Sigma_Y A$ form a basis spanning the columns $B_1$ in $B_1 \Theta'$ of \eqref{linear}, as shown by \citet{gourieroux}. That is, there exists a $r \times r$ nonsingular matrix $M$ such that $B_1 = \Sigma_Y A M$. Therefore, if we let $g_t = A'Y_t$, and let $\alpha^*, B_1^*$ be the parameters of the linear projection of $y_t$ on $(1, g_t')'$ and $u_t^*$ be the resulting projection error such that 

\begin{equation}
y_t = L[y_t \mid (1, g_t)] + u_t^* = \alpha^* + B_1^*g_t + u_t^*
\end{equation}

Then, it implies that the columns of $B_1^*$ are spanned by the columns of $B_1$, that is, $B_1^* = B_1 M^{-1}$. Therefore, we can perform linear regression on the CCA factors, which span the same space as the true factors $f_{1,t}$.

Secondly, the projection errors from the regressions with the CCA factors will not contain systematic measurement errors $v_t$, and are linear functions of the errors $u_t$ and the missing factors $f_{2,t}$ only. However, even so, the projection errors $u_t^*$ may still be mutually correlated, causing the first principal component of the regression residuals to be significant. Therefore, this may not imply that there exists missing factors. However, \citet{ahn} argue that if the first principal component of the residuals only has moderate explanatory power for all response variables or strong explanatory power for a few response variables, then there is no missing factor $f_{2,t}$. However, if the first principal component of the residuals has strong explanatory power for \textbf{all }response variables, then there exists a missing factor not captured by the proxies.

After constructing the CCA factors, we then adopt methodologies from \citet{hair}, using standard statistical tests such as Wilks' Lambda and the Redundancy Index to estimate the number of factors, $r$.

\section{Data} \label{sec:data}

\subsection{P2P Credit Spreads} \label{sec:p2pcs}
Our initial data set consists of 887,440 observations of individual loan data from Lending Club, starting from 1 June, 2007 to 1 December, 2015. Each observation includes features such as the interest rate at which the loan was made, the loan grade and the loan term. We then aggregate individual observations by averaging the interest rates across two cross sections --- loan term and grade, and across monthly periods. There are two loan terms available for P2P loans --- 36 months and 60 months, and 6 grades assigned to each loan --- from A to F.

For 36 month loans, we have 103 monthly aggregated interest rate observations for each grade type, from 1 June, 2007 to 1 December, 2015. However, since 60 month loans were only offered later on, we have 67 aggregated interest rate observations for each grade type, from 1 May, 2010 to 1 Dec, 2015. Table~\ref{tab:cslevel} in the Appendix presents summary statistics for the aggregated interest rates. We then construct the levels of monthly credit spreads by subtracting the yield to maturity of a government bond of the same maturity term (36 months and 60 months respectively) from the P2P interest rate, at each point in time.

Since many empirical studies have found non-stationarity in the levels of credit spreads, we test for stationarity using the Augmented Dicker Fuller (ADF) test and present our findings in Table~\ref{tab:adflevel}. We find that we cannot reject the null hypothesis of a unit root present in an autoregressive model for any of the credit spread levels. Hence, the levels of credit spreads are non-stationary. Therefore, we use first differences in the levels of credit spreads, which represent the change in the levels of the credit spreads from one period to another. We present summary statistics of the differentiated series in Table~\ref{tab:csleveldiff}, and ADF test results for the differentiated series in Table~\ref{tab:adfdiff}. Since none of the $p$-values are above 0.05, we can assume that all the differentiated series are stationary.

Since we will be using the differentiated series, it is necessary to check for cointegration, as co-integrated differenced credit spreads would require an error correction factor model, as in \citet{tu}. Co-integration of a set of time series variables implies that a linear combination of that set of variables integrated of order 1 (first-differenced) is integrated of order 0. Intuitively, economic time series tend to be cointegrated as they are dominated by smooth, long-run equilibrium trends. In our case, there is a possibility of a long-run relation in the levels of P2P credit spreads of different grade quality since it has been found in corporate credit spreads.

In order to test for this, we use the Johansen test, which tests for more than one cointegrating relationship of order 1. In particular, we use the Johansen test with trace, which tests the null hypothesis that the number of cointegration vectors or linear combination is $r = r^* < k$ versus the alternative that $r = k$. We present our findings in Table~\ref{tab:johansen}, and conclude that none of the cointegration relations are statistically significant to reject the null hypothesis. Therefore, we do not include any error-correction mechanisms in our regressions.

\subsection{Macroeconomic Variables} \label{sec:macrovariables}

We use a set of macroeconomic variables that we posit may be important in explaining systematic variation in P2P credit spreads, and present their summary statistics in Table~\ref{tab:macros}.

\begin{description}

\item[General economy.] Since little research exists to find correlation between aggregated P2P loan rates and the general economy, we rely mostly on economic intuition to choose economic variables that reflect the health of the economy and may affect P2P loan rates. Given that \citet{dietrich} found that higher unemployment rates were correlated with higher individual P2P lending rates, we include the U.S.\ seasonally adjusted unemployment rate (UNRATE). However, since the unemployment rate is a quarterly economic indicator, we linearly interpolate between the quarterly observations to obtain monthly observations. Although \citet{okun} observed that the unemployment rate is empirically strongly correlated with gross domestic product (GDP), we choose to include GDP since the relationship between unemployment rate and GDP may have been weakened, especially given the evidence of stagflation. We also choose to include the consumer price index (CPI) and household service payments as a percent of disposable personal income (Debt) in our analysis. A higher unemployment rate may increase the likelihood of lower grade loan default as jobs and income are lost; a higher inflation rate reduces real household income and may increase the likelihood of default; higher proportion of debt repayments for standard loans like housing loans would mean less income for other non-standard loan repayments, like P2P loans, resulting in higher likelihood of P2P loan default.

\item[Government bond market.] P2P loans can be seen as credit investments, except with a higher rate of default since corporations are more likely to repay loans due to regulations. Given their similarity, P2P credit spreads are likely to be correlated with the risk-free government bond interest rates, as \citet{duffee} found in corporate credit spreads and \citet{dietrich} found in individual P2P loans. The risk-free government bond interest rates form the Treasury yield curve, a cross-section of Treasuries across different maturities at a single point in time, which has been found to contain three factors --- level, slope and curvature. To account for the level factor, we use the first-differenced interest rate of the 10-year Treasury bill ($\Delta$ 10Y-rf). We also include the slope factor, as \citet{litterman} have documented, by taking the difference between the 10-year Treasury bill rate and the 1-year Treasury bill rate (rf-slope). The curvature factor has little meaningful interpretation at present, therefore, we do not include it in our analysis. Our Treasury yields data and general economic variables data are obtained from the Federal Reserve Bank of St.\ Louis' FRED database.

\item[Equity market.] P2P loans may be affected by the equity market as the rates of return on P2P loans are comparable to equities, resulting in substitutability between P2P loans and equities as investments, especially in periods of low interest rates. We use observations from the S\&P Index (SPX) compiled by Bloomberg to represent the general performance of the equity market. We also include the first differenced VIX index ($\Delta$VIX), which tracks changes in the market's expectations of 30-day equity volatility. A large change in the VIX index implies investor uncertainty, which can reduce investor demand in investing in P2P loans, resulting in higher P2P interest rates. We are also interested in analyzing whether P2P credit spreads may be related to \citet{fama} three factors in stock returns. In particular, the equity premium from the small-minus-big (SMB) factor or the high-minus-low (HML) factor may be correlated to P2P credit spreads. If the equity premium on HML and SMB factors are low, investors may look for other sources of return, turning to P2P loans and reducing P2P interest rates. We obtain both factor returns from French's publicly available data library website. 

\end{description}

\section{Results} \label{sec:results}

\subsection{Multiple Regressions} \label{sec:multregress}
We first analyze our data using multiple regressions to establish a benchmark case. The specification for our full regression model is given by 
\begin{equation} \label{regression}
\begin{split}
\Delta \operatorname{CS}_{it} = & \beta_{0,i} + \beta_{1,i} \operatorname{UNRATE}_t + \beta_{2,i} \operatorname{GDP}_t + \beta_{3,i} \operatorname{CPI}_t + \beta_{4,i} \operatorname{Debt}_t + \beta_{5,i}\Delta \operatorname{10Yrf}_t + \beta_{5,i} \operatorname{Slope}_t \\ & + \beta_{6,i} \operatorname{SPX}_t + \beta_{7,i} \Delta \operatorname{VIX}_t + \beta_{8,i} \operatorname{SMB}_t + \beta_{9,i} \operatorname{HML}_t + e_{it}
\end{split}
\end{equation}

We perform our regressions over two panels --- one with the loan grade cross-section, and the other with the loan term cross-section. For example, for the grade-A loan regressions, we combine data of both 36-month grade-A loan observations and 60-month grade-A loan observations; for the 36-month loan regressions, we combine data of 36-month loan observations across all the grades. We present our results in Table~\ref{tab:olsfull} and Table~\ref{tab:olsfull2} of the Appendix.

Similar to corporate credit spreads, the results show that changes in the risk-free rate are statistically and economically significant in explaining changes in P2P credit spreads across all grade types and terms. The magnitudes of the coefficients are also similar for all the regressions, and are negative, comparable to the coefficients found in corporate credit spreads in \citet{ahn}. This implies that an increase in the 10-year Treasury yield is associated with a decrease in P2P loan rates across all grade and term types.

Focusing first on Panel A, we find that the unemployment rate is a significant and negative predictor for higher grade credit spreads. Although the equity index is also statistically significant predictor for higher grade credit spreads, the coefficient is close to zero, and thus, not economically meaningful. We also find that HML and SMB are statistically significant for higher grade credit spreads, with the HML factor coefficient being positive and the SMB coefficient being negative.

For Panel B, unemployment rate remains a significant and negative predictor for both term types. Likewise, the SPX predictor, while significant, is not economically meaningful. The HML and SMB factor predictors are statistically significant for both term types, and also in the same sign direction as our findings in Panel A. Interestingly, the change in VIX is a significant and positive predictor for shorter term loans.

Since the predictors in the fully specified model are likely to be correlated by definition for factor construction, we may face problems of multicollinearity resulting in large standard errors and thus, conservative hypothesis testing. Therefore, we use AIC stepwise forward-backward regression for variable selection, shown in Tables~\ref{tab:olsstatistics} and \ref{tab:olsstatistics2}.

Our conclusions from the AIC stepwise regressions are not materially different from what we found with the full specification regressions. However, we note that for lower grade credit spreads, CPI is chosen as a significant predictor instead of the unemployment rate. This is not surprising given the high correlation between CPI and unemployment rate, as documented in Table~\ref{tab:corrmacro}.

Next, we examine the estimated residuals using principal components to ascertain the existence of a strong common component in the residuals. We follow \citet{ahn} and extract the first principal component from the residuals, before adding the extracted series into the OLS regression. We present our results in Tables~\ref{tab:ols1pc} and \ref{tab:ols1pc2} in the Appendix. We see that similar to corporate credit spreads, the change in adjusted $R^2$ increases sharply after adding in the residual principal component. The average increase in adjusted $R^2$ for grade type regressions is 0.440 while the average increase for term type regressions is 0.536. 

\subsection{Canonical Correlation Analysis} \label{sec:ccaresult}
Following the CCA technique we laid out in Section~\ref{sec:ccafactor}, we then let the first set of variables $Y$ be the set of all P2P credit spreads across grade types and loan terms. Therefore, we have twelve variables in set $Y$. Then, we let the second set of variables $Z$ be the ten macroeconomic proxy variables found in \eqref{regression}. We first report the canonical correlation coefficients and the eigenvalues of the canonical roots in Table~\ref{tab:ccaeigen}, following the same analysis procedure in \citet{hair}. A quick analysis of the eigenvalues would suggest that the optimal number of factors would be five, as the eigenvalues from the fifth onward seem to level off.

To test the extent of correlation between the two sets of variables, we use Wilk's lambda in Table~\ref{tab:lambdatest}, which tests sequential hypotheses that the $k$th canonical variate and all that follow it are zero. The results show that at 5\% significance level, we can reject the null hypothesis $H_0: \rho_1 = \rho_2 = \dots = \rho_6 = 0$ where $\rho_k = \operatorname{Corr}(U_k, V_k)$.

This would suggest that having six factors would be optimal in capturing systematic variation in both sets of variables. However, while the squared canonical correlations (roots) give an estimate of the shared variance between the canonical variates, i.e., the variance shared by linear combinations of the two sets of variables, they do not capture the proportion of variance of the respective sets of variables. The redundancy index provides a measure of the extent to which a set of variables (taken as a set) explains variation in the other set of variables (taken one at a time), by computing the proportion of variances of the individual variables in each set which are accounted for by all the variables in the other set through the canonical variates. We report the redundancies for P2P credit spread variables in Table~\ref{tab:redunindices}.

The redundancy analysis suggests that using the first three canonical variates is sufficient, since all the other canonical variates have very low redundancy indices. We note that the number of canonical variates found by the redundancy index is the same as the number of canonical variates found by \citet{ahn} by GMM estimation for corporate spreads.

Before analyzing and interpreting each factor, we compute canonical cross-loadings, which are the pairwise correlations between the three canonical variates and the set of macroeconomic variables, in Table~\ref{tab:pairwisecorr}. We will then use the canonical cross-loadings along with the CCA factor regression estimates to interpret the factors. 

\subsection{CCA Factor Regressions} \label{sec:ccafactorresult}

After extracting three factors by estimating the first three canonical correlations, we estimate our CCA factor model with the following specification, where $\operatorname{Factor1}$ is the vector of first canonical variate $Y$ scores, $\operatorname{Factor2}$ is the vector of second canonical variate $Y$ scores and so on. 
\begin{equation} \label{ccaregression}
\Delta \operatorname{CS}_{it} = \beta_{0,i} + \beta_{1,i} \operatorname{Factor1}_t + \beta_{2,i} \operatorname{Factor2}_t + \beta_{3,i} \operatorname{Factor3}_t + u_t^*
\end{equation}
The estimation results in Tables~\ref{tab:ccafactor} and \ref{tab:ccafactor2} show a large increase in explanatory power for grade type regressions, compared to the regressions in Section~\ref{sec:multregress}. The adjusted $R^2$ values are all greater than 0.7, while none of the loan type regressions in Section~\ref{sec:multregress} had an adjusted $R^2$ greater than $0.5$. Furthermore, the adjusted $R^2$ increases as the loan type becomes lower, that is, as the loan becomes less credit-worthy. We observe that the coefficient for $\operatorname{Factor1}$ increases for loans that are less credit-worthy, while the coefficient for $\operatorname{Factor2}$ is not monotonic --- becoming more negative before becoming less negative and ultimately switching sign for the last category F. The coefficients for $\operatorname{Factor3}$ do not follow a trend, but is lowest for category A and highest for category E.

On the other hand, when loans are sorted by term type, we find that the adjusted $R^2$ values drop sharply, implying that the factors explain little of the systematic variation of interest rates in each term category. Furthermore, $\operatorname{Factor2}$ is not statistically significant in the regressions for both term types. The coefficients for both $\operatorname{Factor1}$ and $\operatorname{Factor3}$ are statistically significant and positive, and increase for the higher term category of 60 months.

The first canonical variate is strongly negatively correlated with the risk-free Treasury slope (rf-slope), household debt (Debt), and unemployment rate (UNRATE), but strongly positively correlated with inflation (CPI). This suggests that $\operatorname{Factor1}$ may represent a macro non-default factor. In periods of economic expansion, the unemployment rate decreases while inflation increases, as predicted by the Phillips curve (1958). Due to increased household income as a result of lower unemployment rates, household service payments as a percentage of disposable income also decreases, assuming constant household debt levels. However, the macro non-default factor is negatively correlated with the slope of the risk-free rate, which represents investors' expectations about future risk-free rates. In particular, a decreased slope implies that investors do not expect short-term risk-free rates to rise. Prior to quantitative easing in 2008, this typically implies that investors are bearish about the economy in the short run. However, given the general low interest rate environment from 2008 to 2015, the traditionally established link between the Treasury yields slope and macroeconomic expectations can be questioned, and thus, will not affect our interpretation of $\operatorname{Factor1}$ as a macro non-default factor. Instead, this adds some subtlety in our interpretation of the non-default factor --- although the economy is improving, economic perceptions are still bearish.

Since $\operatorname{Factor1}$ has a positive coefficient in the CCA factor regressions, this implies that an decrease in the likelihood of macro default is correlated with a decrease in P2P credit spreads across all grade types and term types. This is because if the probability of default across the economy decreases, the probability of default for P2P loans will also decrease, reducing the P2P interest rates. Furthermore, the magnitude of the effect increases for lower grade P2P loans, that is, loans that are less credit-worthy, as well as for longer term P2P loans. This may be explained by the fact that recessionary economic conditions are likely to affect lower grade P2P debtors more, since lower grade P2P loans are correlated with worse credit history or borrow-specific attributes like lower income levels. The effect is larger for longer term loans as well since longer term P2P loans have a higher risk of default as long term economic horizons are more difficult to predict. Therefore, the effect of macro-default on P2P loans is magnified, further increasing the interest rates of longer term loans.

The most apparent difference between the first canonical variate and the second canonical variate is that the risk-free Treasury slope changes from a negative correlation to a positive correlation, and GDP changes from a slight negative correlation to a strong positive correlation. In addition, equity volatility is slightly positively correlated with the second canonical variate, albeit at a small magnitude. This factor can be interpreted as latent market uncertainty that may be correlated with short-term market rallies. Given the low interest rate environment mentioned earlier, market uncertainty may be correlated with higher slopes i.e., a steeper Treasuries yield curve, as investors demand a higher premium for investing in long-term government bonds. At the same time, within the time frame of our dataset, US GDP experienced two sudden swings (a sudden drop followed by a sudden increase), although the unemployment rate steadily decreased. Therefore, given that GDP can function like a signal rather than an indicator in the economy, sudden swings in GDP may thus be correlated with greater uncertainty. This hypothesis is directly supported by the fact that the change in equity volatility is positively correlated with $\operatorname{Factor2}$.

Our CCA factor regressions reveal that $\operatorname{Factor2}$ has a negative coefficient, implying that an increase in latent traditional market uncertainty is associated with a decrease in credit spreads for all grade types except grade F. The negative coefficient, however, is strongest for grade type B, but consistently decreases from B to E. We first explain the negative coefficient, which may arise as uncertainty about traditional investment products encourages more investors to seek higher returns through other means, such as P2P platforms, increasing money supply for the P2P market. This causes P2P interest rates to fall and the P2P credit spread to decrease. Next, we explain why market uncertainty has a reduced effect on the credit spread of lower grade loan categories. By the same argument, investors who wish to diversify their portfolio and switch from more traditional products to P2P lending will seek P2P investments that are more substitutable with traditional products in terms of the default risk they entail. Since P2P loans have a higher default risk than corporate bonds, we can expect P2P loans of grade A or B to be substitutable with corporate bonds of grade B or C. Therefore, demand for lower grade P2P loans do not increase proportionally. On the other hand, our CCA factor regressions by loan term type suggest that latent macroeconomic uncertainty have no significant effect on interest rates defined by their term type. This may make sense as market uncertainty will affect P2P interest rates regardless of term type, given the range of duration risk appetites of investors who substitute traditional investments with P2P investments, resulting in little difference in credit spreads across term types.

Interpretation for $\operatorname{Factor3}$ is less apparent as the third canonical variate is not strongly correlated with any of our macroeconomic proxies. Instead, we observe that $\operatorname{Factor3}$ has a stronger positive correlation with the HML equity factor, and a stronger negative correlation with the SMB equity factor compared to the first two P2P factors. Therefore, if necessary, we are inclined to interpret the third canonical variate as the fundamental value of the equity market. This is expected to be positively correlated to the HML factor, which is the spread in returns between value and growth stocks. The fundamental value of the equity market can also be negatively correlated with the SMB factor, since the SMB factor is comprised of volatile small-cap stocks with returns that are not associated with their fundamental value. Furthermore, \citet{berg} find that HML and SMB factors are not sensitive to macroeconomic risks, which can expain why $\operatorname{Factor3}$ has relatively low correlation with the other macroeconomic proxies, as compared to $\operatorname{Factor1}$ or $\operatorname{Factor2}$. The CCA factor regressions show that an increase in $\operatorname{Factor3}$ is associated with an increase in P2P credit spreads across all grade types. A similar relationship is found by \citet{vas}, who concluded that an increase in the returns of HML is associated with an  increase in the difference between the return on long-term government and corporate bonds. However, our interpretation of $\operatorname{Factor3}$ is tenuous, as the correlations with the macroeconomic proxies are weak.

We note that while the level of the yield curve ($\Delta$ 10Y-rf) is statistically significant in our fully specified multiple regressions, it has very low correlation with the factors extracted from the canonical correlations. Given this disparity, we are inclined to think that OLS regressions face issues of multicollinearity, as the level of the yield curve is heavily correlated with the other macroeconomic proxies, like inflation rate. Therefore, CCA factors allow us to extract more meaningful and accurate insights about systematic variations in P2P spreads. However, while our interpretation are supported by economic theory and intuition, we concede that they are tentative.

As before, we extract the first principal component from the residuals and include the extracted series in our CCA factor regressions, see Tables~\ref{tab:ccaregression} and \ref{tab:ccaregression2}. We find that the average increase in adjusted $R^2$ values for grade type regressions is 0.103, which is significantly smaller compared to the initial OLS regressions. On the other hand, the average increase in adjusted $R^2$ values for term type regressions is 0.544, which is similar to the average increase in the OLS regressions. Therefore, it appears that the CCA factors do account for most of the common variation in P2P spreads sorted by grade type, and that the unexplained erros are idiosyncratic, implying no missing factors $f_{2,t}$ in \eqref{factor}. On the other hand, the CCA factors fail to account for common variation in P2P spreads sorted by term type, implying that there are missing factors $f_{2,t}$. Therefore, we are inclined to believe that macroeconomic conditions affect common variation in the interest rates of different P2P grades, but of different P2P terms.  

\section{Conclusion} \label{sec:conclusion}
We use canonical correlations to extract latent macroeconomic factors, in order to analyze the effect that the wider economy has on aggregated P2P interest rates. The technique also offers the advantage of allowing us to accurately verify whether there are missing non-macroeconomic factors that affect common variation in P2P credit spreads, by eliminating systematic measurement errors in macroeconomic variables that can affect our results. We show that the common variation in the grade structure of credit spreads is explained by three factors --- macro non-default, market uncertainty and (possibly) fundamental market value.

\section*{Acknowledgments}

The work in this article is generously supported by DARPA D15AP00109 and NSF IIS 1546413. The authors are grateful to the Stevanovich Center at the University of Chicago for permission to use its facilities.

\clearpage

\section*{Appendix}

\begin{table}[ht]
	\centering
	\begin{tabular}{rrrrr}
		\hline
		& min & max & mean & sd \\ 
		\hline
		36-A & 2.82 & 7.96 & 6.37 & 1.03 \\ 
		36-B & 4.47 & 12.04 & 9.72 & 1.73 \\ 
		36-C & 5.85 & 15.31 & 12.30 & 2.26 \\ 
		36-D & 7.61 & 18.34 & 14.72 & 2.80 \\ 
		36-E & 9.03 & 21.58 & 16.92 & 3.32 \\ 
		36-F & 10.56 & 23.70 & 19.43 & 3.82 \\ 
		60-A & 4.89 & 8.29 & 6.87 & 0.86 \\ 
		60-B & 8.23 & 11.88 & 9.99 & 1.18 \\ 
		60-C & 10.51 & 15.52 & 13.03 & 1.34 \\ 
		60-D & 12.97 & 18.51 & 15.99 & 1.55 \\ 
		60-E & 14.67 & 20.99 & 18.46 & 1.90 \\ 
		60-F & 16.30 & 23.18 & 21.19 & 2.17 \\ 
		\hline
	\end{tabular}
	\caption{Description statistics: levels of monthly credit spreads across 2 maturity terms and 6 grades} \label{tab:cslevel}
\end{table}

\begin{table}[ht]
	\centering
	\begin{tabular}{rrr}
		\hline
		& Test Statistic & $p$-value \\ 
		\hline
		36-A & $-1.73$ & 0.69 \\ 
		36-B & $-1.94$ & 0.60 \\ 
		36-C & $-1.30$ & 0.87 \\ 
		36-D & $-0.33$ & 0.99 \\ 
		36-E & $-0.48$ & 0.98 \\ 
		36-F & $-0.48$ & 0.98 \\ 
		60-A & $-1.62$ & 0.74 \\ 
		60-B & $-1.70$ & 0.70 \\ 
		60-C & $-1.17$ & 0.91 \\ 
		60-D & $-1.01$ & 0.93 \\ 
		60-E & $-1.06$ & 0.92 \\ 
		60-F & $-0.48$ & 0.98 \\ 
		\hline
	\end{tabular}
\caption{ADF test statistics and $p$-values for levels of monthly credit spreads} \label{tab:adflevel}
\end{table}

\begin{table}[ht]
	\centering
	\begin{tabular}{rrrrr}
		\hline
		& Min & Max & Mean & S.D. \\ 
		\hline
		36-A & $-0.85$ & 1.17 & 0.02 & 0.30 \\ 
		36-B & $-0.85$ & 1.14 & 0.03 & 0.35 \\ 
		36-C & $-0.69$ & 1.04 & 0.05 & 0.33 \\ 
		36-D & $-0.98$ & 1.41 & 0.07 & 0.36 \\ 
		36-E & $-1.53$ & 1.98 & 0.09 & 0.44 \\ 
		36-F & $-0.93$ & 2.71 & 0.11 & 0.49 \\ 
		60-A & $-1.05$ & 0.99 & 0.01 & 0.30 \\ 
		60-B & $-0.77$ & 0.89 & $-0.02$ & 0.34 \\ 
		60-C & $-0.97$ & 0.83 & 0.00 & 0.31 \\ 
		60-D & $-0.80$ & 1.47 & 0.02 & 0.35 \\ 
		60-E & $-0.53$ & 1.66 & 0.03 & 0.36 \\ 
		60-F & $-0.47$ & 2.03 & 0.08 & 0.38 \\ 
		\hline
	\end{tabular}
\caption{Description statistics: first differences of monthly credit spreads across 2 maturity terms and 6 grades} \label{tab:csleveldiff}
\end{table}

\begin{table}[ht]
	\centering
	\begin{tabular}{rrr}
		\hline
		& Test Statistic & $p$-value \\ 
		\hline
		36-A & $-7.53$ & 0.01 \\ 
		36-B & $-7.02$ & 0.01 \\ 
		36-C & $-7.23$ & 0.01 \\ 
		36-D & $-8.20$ & 0.01 \\ 
		36-E & $-9.68$ & 0.01 \\ 
		36-F & $-10.70$ & 0.01 \\ 
		60-A & $-8.65$ & 0.01 \\ 
		60-B & $-6.38$ & 0.01 \\ 
		60-C & $-7.03$ & 0.01 \\ 
		60-D & $-6.30$ & 0.01 \\ 
		60-E & $-6.24$ & 0.01 \\ 
		60-F & $-7.50$ & 0.01 \\ 
		\hline
	\end{tabular}
\caption{ADF test statistics and $p$-values for first differences of monthly credit spreads} \label{tab:adfdiff}
\end{table}

\begin{table}[ht]
	\centering
	\begin{tabular}{rrrr}
		\hline
		& Test Statistic & Critical Value (5\%) \\ 
		\hline
		$r \leq 5$  & 2.79 & 8.18 \\ 
		$r \leq 4$  & 9.06  & 17.95 \\ 
		$r \leq 3$  & 16.49 & 31.52 \\ 
		$r \leq 2$  & 27.74 & 48.28 \\ 
		$r \leq 1$  & 48.82 & 70.60 \\ 
		$r = 0$  & 74.57 & 90.39 \\ 
		\hline
		$r \leq 5$ & 4.30 & 8.18 \\ 
		$r \leq 4$ & 11.08  & 17.95 \\ 
		$r \leq 3$  & 21.01 & 31.52 \\ 
		$r \leq 2$  & 34.40 & 48.28 \\ 
		$r \leq 1$  & 51.69 & 70.60 \\ 
		$r = 0$  & 77.46 & 90.39 \\ 
		\hline
	\end{tabular}
	\caption{Johansen test statistics and critical values for levels of monthly credit spreads} \label{tab:johansen}
\end{table}

\begin{table}[ht]
	\centering
	\begin{tabular}{rrrrr}
		\hline
		& Min & Max & Mean & S.D. \\ 
		\hline
		UNRATE & 4.60 & 10.00 & 7.37 & 1.73 \\ 
		GDP & $-8.20$ & 5.00 & 1.03 & 2.42 \\
		CPI & 207.23 & 238.15 & 224.96 & 9.63 \\
		Debt & 9.89 & 13.21 & 11.05 & 1.15 \\ 
		$\Delta$10Y-rf & $-1.08$ & 0.64 & $-0.02$ & 0.27 \\ 
		rf-slope & $-0.07$ & 3.43 & 2.11 & 0.76 \\ 
		SPX & 735.09 & 2107.39 & 1449.21 & 365.96 \\ 
		$\Delta$VIX & $-15.28$ & 20.50 & $-0.08$ & 5.57 \\
		SMB & $-4.29$ & 6.11 & 0.13 & 2.27 \\ 
		HML & $-11.25$ & 7.85 & $-0.24$ & 2.73 \\ 
		\hline
	\end{tabular}
\caption{Description statistics: macroeconomic variables of monthly frequency} \label{tab:macros}
\end{table}

\begin{table}[ht]
	\centering
	\begin{tabular}{rrrrrrrrrrr}
		\hline
		& CPI & UNRATE & Debt & GDP & SPX & $\Delta$10Y-rf & rf-slope & SMB & HML & $\Delta$VIX \\ 
		\hline
		CPI & 1.00 & $-0.94$ & $-0.95$ & 0.23 & 0.91 & 0.04 & $-0.34$ & $-0.10$ & 0.09 & 0.09 \\ 
		UNRATE & $-0.94$ & 1.00 & 0.84 & $-0.35$ & $-0.97$ & $-0.04$ & 0.32 & 0.07 & $-0.03$ & $-0.06$ \\ 
		Debt & $-0.95$ & 0.84 & 1.00 & $-0.07$ & $-0.79$ & $-0.04$ & 0.51 & 0.11 & $-0.18$ & $-0.09$ \\ 
		GDP & 0.23 & $-0.35$ & $-0.07$ & 1.00 & 0.40 & $-0.02$ & 0.19 & 0.01 & $-0.33$ & 0.06 \\ 
		SPX & 0.91 & $-0.97$ & $-0.79$ & 0.40 & 1.00 & 0.07 & $-0.16$ & $-0.04$ & $-0.02$ & 0.03 \\ 
		$\Delta$10Y-rf & 0.04 & $-0.04$ & $-0.04$ & $-0.02$ & 0.07 & 1.00 & 0.12 & 0.31 & 0.44 & $-0.29$ \\ 
		rf-slope & $-0.34$ & 0.32 & 0.51 & 0.19 & $-0.16$ & 0.12 & 1.00 & 0.04 & $-0.10$ & $-0.04$ \\ 
		SMB & $-0.10$ & 0.07 & 0.11 & 0.01 & $-0.04$ & 0.31 & 0.04 & 1.00 & $-0.01$ & $-0.15$ \\ 
		HML & 0.09 & $-0.03$ & $-0.18$ & $-0.33$ & $-0.02$ & 0.44 & $-0.10$ & $-0.01$ & 1.00 & 0.04 \\ 
		$\Delta$VIX & 0.09 & $-0.06$ & $-0.09$ & 0.06 & 0.03 & $-0.29$ & $-0.04$ & $-0.15$ & 0.04 & 1.00 \\ 
		\hline
	\end{tabular}\caption{Description statistics: correlations between macroeconomic variables} \label{tab:corrmacro}
\end{table}

\begin{landscape}
\begin{adjustwidth*}{-1cm}{-1cm}
	\renewcommand{\arraystretch}{1.3}
	\begin{table}[ht]
		\centering
		\begin{tabular}{rrrrrrrrrrrrr}
			\hline
			& (Intercept) & UNRATE & GDP & CPI & Debt & $\Delta$10Yrf & Slope & SPX & $\Delta$VIX & HML & SMB & Adj.\ $R^2$ \\ 
			\hline
			A & 6.741 & $\mathbf{-0.014}$ & 0.006 & $-0.015$ & $-0.159$ & $\mathbf{-0.614}$ & 0.073 & $\mathbf{-0.001}$ & $-0.001$ & $\mathbf{0.014}$ & $-0.015$ & 0.339 \\ 
			& 1.546 & $-2.210$ & 0.409 & $-1.150$ & $-1.395$ & $-7.270$ & 1.359 & $-1.864$ & $-0.166$ & 1.681 & $-1.625$ &  \\ 
			B & 6.795 & $\mathbf{-0.159}$ & $-0.000$ & $-0.014$ & $-0.156$ & $\mathbf{-0.718}$ & 0.096 & $\mathbf{-0.001}$ & 0.002 & 0.013 & $\mathbf{-0.020}$ & 0.400 \\ 
			& 1.415 & $-2.275$ & $-0.018$ & $-0.982$ & $-1.241$ & $-7.742$ & 1.628 & $-2.106$ & 0.431 & 1.488 & $-2.006$ &  \\ 
			C & 1.434 & $-0.023$ & $-0.007$ & $-0.003$ & $-0.004$ & $\mathbf{-0.749}$ & $-0.041$ & $-0.000$ & 0.003 & 0.009 & $-0.006$ & 0.479 \\ 
			& 0.345 & $-0.387$ & $-0.535$ & $-0.271$ & $-0.035$ & $-9.335$ & $-0.807$ & $-0.850$ & 0.902 & 1.193 & $-0.725$ &  \\ 
			D & 3.785 & $-0.019$ & $-0.008$ & $-0.012$ & $-0.057$ & $\mathbf{-0.745}$ & $-0.030$ & $-0.000$ & 0.002 & 0.005 & $-0.004$ & 0.361 \\ 
			& 0.746 & $-0.260$ & $-0.504$ & $-0.805$ & $-0.430$ & $-7.608$ & $-0.487$ & $-0.398$ & 0.543 & 0.514 & $-0.393$ &  \\ 
			E & 2.515 & $-0.050$ & 0.009 & $-0.004$ & $-0.045$ & $\mathbf{0.674}$ & $-0.009$ & $-0.000$ & 0.003 & $-0.000$ & $-0.008$ & 0.245 \\ 
			& 0.395 & $-0.536$ & 0.470 & $-0.202$ & $-0.268$ & $-5.479$ & $-0.122$ & $-1.162$ & 0.522 & $-0.016$ & $-0.591$ &  \\ 
			F & 4.884 & $-0.035$ & $-0.017$ & $-0.015$ & $-0.093$ & $\mathbf{-0.756}$ & 0.065 & $-0.000$ & 0.001 & 0.001 & $-0.024$ & 0.223 \\ 
			& 0.683 & $-0.340$ & $-0.784$ & $-0.726$ & $-0.498$ & $-5.509$ & 0.743 & $-0.255$ & 0.169 & 0.076 & $-1.619$ &  \\ 
			\hline
		\end{tabular}
		\caption{Panel A OLS Regressions: Full specification regressions according to Equation \eqref{regression}, across loan grade types} \label{tab:olsfull}
	\end{table}
	\renewcommand{\arraystretch}{1.3}
	\begin{table}[ht]
		\centering
		\begin{tabular}{rrrrrrrrrrrrr}
			\hline
			& (Intercept) & UNRATE & GDP & CPI & Debt & $\Delta$10Yrf & Slope & SPX & $\Delta$VIX & HML & SMB & Adj.\ $R^2$ \\ 
			\hline
			36-m & 4.503 & $\mathbf{-0.081}$ & $-0.001$ & $-0.011$ & $-0.086$ & $\mathbf{-0.764}$ & 0.028 & $\mathbf{-0.000}$ & $\mathbf{0.004}$ & $\mathbf{0.013}$ & $\mathbf{-0.018}$ & 0.358 \\ 
			& 1.759 & $-2.126$ & $-0.061$ & $-1.421$ & $-1.282$ & $-13.206$ & 0.844 & $-2.119$ & 1.795 & 2.733 & $-3.023$ &  \\ 
			60-m & 4.004 & $\mathbf{-0.070}$ & $-0.003$ & $-0.010$ & $-0.069$ & $\mathbf{-0.687}$ & 0.030 & $\mathbf{-0.000}$ & 0.003 & $\mathbf{0.013}$ & $\mathbf{-0.018}$ & 0.337 \\ 
			& 1.429 & $-1.700$ & $-0.304$ & $-1.203$ & $-0.940$ & $-12.387$ & 0.852 & $-1.753$ & 1.192 & 2.446 & $-2.797$ &  \\ 
			\hline
		\end{tabular}
		\caption{Panel B OLS Regressions: Full specification regressions according to Equation \eqref{regression}, across loan term types} \label{tab:olsfull2}
	\end{table}
\end{adjustwidth*}

\renewcommand{\arraystretch}{1.5}
\begin{table}[ht]
	\centering
	\begin{tabular}{rrrrrrrrr}
		\hline
		& (Intercept) & UNRATE & CPI & $\Delta$10Yrf & SPX & HML & SMB & Adj.\ $R^2$ \\ 
		\hline
		A & 0.84 & $-0.05$ &  & $-0.60$ & $-0.00$ & 0.01 & $-0.02$ & 0.35 \\ 
		& 3.31 & $-2.89$ &  & $-7.31$ & $-3.32$ & 1.56 & $-1.86$ &  \\ 
		B & 1.14 & $-0.07$ &  & $-0.70$ & $-0.00$ & 0.01 & $-0.02$ & 0.41 \\ 
		& 4.09 & $-3.32$ &  & $-7.76$ & $-4.43$ & 1.42 & $-2.33$ &  \\ 
		C & 1.04 & $-0.05$ &  & $-0.77$ & $-0.00$ &  &  & 0.49 \\ 
		& 4.32 & $-3.06$ &  & $-10.46$ & $-5.08$ &  &  &  \\ 
		D & 2.25 &  & $-0.01$ & $-0.78$ &  &  &  & 0.38 \\ 
		& 3.94 &  & $-3.87$ & $-8.83$ &  &  &  &  \\ 
		E & 0.48 &  &  & $-0.73$ & $-0.00$ &  &  & 0.27 \\ 
		& 4.08 &  &  & $-6.60$ & $-3.73$ &  &  &  \\ 
		F & 2.20 &  & $-0.01$ & $-0.74$ &  &  & $-0.03$ & 0.25 \\ 
		& 2.70 &  & $-2.59$ & $-5.72$ &  &  & $-1.74$ &  \\ 
		\hline
	\end{tabular}
	\caption{Panel A OLS Regressions: AIC stepwise regressions, across loan term types} \label{tab:olsstatistics}
\end{table}

\begin{table}[ht]
	\centering
	\begin{tabular}{rrrrrrrrr}
		\hline
		& (Intercept) & UNRATE & $\Delta$10Yrf & SPX & $\Delta$VIX & HML & SMB & Adj.\ $R^2$ \\ 
		\hline
		36-m & 0.97 & $-0.05$ & $-0.67$ & $-0.00$ & 0.00 & 0.01 & $-0.02$ & 0.36 \\ 
		& 6.91 & $-4.84$ & $-13.41$ & $-7.43$ & 1.84 & 2.55 & $-3.14$ &  \\ 
		60-m & 1.01 & $-0.05$ & $-0.68$ & $-0.00$ &  & 0.01 & $-0.02$ & 0.34 \\ 
		& 6.51 & $-4.46$ & $-12.54$ & $-7.16$ &  & 2.33 & $-3.11$ &  \\ 
		\hline
	\end{tabular}
	\caption{Panel B OLS Regressions: AIC stepwise regressions, across loan term types} \label{tab:olsstatistics2}
\end{table}

\begin{table}[ht]
	\centering
	\begin{tabular}{rrrrrrrrrrrrrr}
		\hline
		& (Intercept) & UNRATE & GDP & CPI & Debt &  $\Delta$10Yrf & Slope & SPX & $\Delta$VIX & HML & SMB & PC1 &  Adj.\ $R^2$ \\ 
		\hline
		A & 6.94 & $-0.14$ & 0.00 & $-0.02$ & $-0.16$ & $-0.61$ & 0.07 & $-0.00$ & $-0.00$ & 0.01 & $-0.01$ & $-0.09$ & 0.70 \\ 
		 & 2.31 & $-3.14$ & 0.48 & $-1.78$ & $-2.06$ & $-10.65$ & 1.76 & $-2.56$ & $-0.17$ & 2.50 & $-2.36$ & $-13.17$ &  \\ 
		B & 7.08 & $-0.15$ & $-0.00$ & $-0.01$ & $-0.16$ & $-0.71$ & 0.09 & $-0.00$ & 0.00 & 0.01 & $-0.02$ & $-0.11$ & 0.81 \\ 
		 & 2.58 & $-3.89$ & $-0.20$ & $-1.88$ & $-2.23$ & $-13.53$ & 2.61 & $-3.48$ & 0.83 & 2.63 & $-3.66$ & $-17.71$ &  \\ 
		C & 2.31 & $-0.03$ & $-0.01$ & $-0.01$ & $-0.02$ & $-0.74$ & $-0.05$ & $-0.00$ & 0.00 & 0.01 & $-0.01$ & $-0.10$ & 0.85 \\ 
		 & 1.05 & $-0.79$ & $-0.91$ & $-1.00$ & $-0.33$ & $-17.34$ & $-1.75$ & $-1.40$ & 1.91 & 2.31 & $-1.30$ & $-19.68$ &  \\ 
		D & 3.53 & $-0.03$ & $-0.01$ & $-0.01$ & $-0.06$ & $-0.74$ & $-0.02$ & $-0.00$ & 0.00 & 0.01 & $-0.00$ & $-0.12$ & 0.87 \\ 
		 & 1.56 & $-0.81$ & $-0.86$ & $-1.56$ & $-0.95$ & $-17.12$ & $-0.69$ & $-1.22$ & 1.11 & 1.14 & $-0.84$ & $-24.79$ &  \\ 
		E & 1.02 & $-0.05$ & 0.01 & 0.00 & $-0.02$ & $-0.69$ & 0.01 & $-0.00$ & 0.00 & $-0.00$ & $-0.01$ & $-0.15$ & 0.80 \\ 
		 & 0.31 & $-1.07$ & 0.92 & 0.21 & $-0.24$ & $-10.87$ & 0.16 & $-2.57$ & 0.77 & $-0.07$ & $-1.18$ & $-20.30$ &  \\ 
		F & 4.75 & $-0.03$ & $-0.02$ & $-0.01$ & $-0.09$ & $-0.76$ & 0.06 & $-0.00$ & 0.00 & 0.00 & $-0.02$ & $-0.14$ & 0.65 \\ 
		 & 0.99 & $-0.47$ & $-1.21$ & $-1.06$ & $-0.72$ & $-8.26$ & 1.07 & $-0.36$ & 0.25 & 0.12 & $-2.43$ & $-13.68$ &  \\ 
		\hline
	\end{tabular}\caption{Panel A PC OLS Regressions: Full specification regressions according to Equation \eqref{regression} including first PC of residuals, across loan term types} \label{tab:ols1pc}
\end{table}

\begin{table}[ht]
	\centering
	\begin{tabular}{rrrrrrrrrrrrrr}
		\hline
		& (Intercept) & UNRATE & GDP & CPI & Debt &  $\Delta$10Yrf & Slope & SPX & $\Delta$VIX & HML & SMB & PC1 &  Adj.\ $R^2$ \\ 
		\hline
		36-m & 4.34 & $-0.08$ & $-0.00$ & $-0.01$ & $-0.08$ & $-0.67$ & 0.03 & $-0.00$ & 0.00 & 0.01 & $-0.02$ & 0.21 & 0.88 \\ 
		 & 3.86 & $-4.87$ & $-0.23$ & $-3.02$ & $-2.88$ & $-30.11$ & 1.87 & $-5.05$ & 3.91 & 5.97 & $-6.65$ & 49.97 &  \\ 
		60-m & 4.17 & $-0.07$ & $-0.00$ & $-0.01$ & $-0.07$ & $-0.69$ & 0.03 & $-0.00$ & 0.00 & 0.01 & $-0.02$ & 0.22 & 0.88 \\ 
		 & 3.53 & $-4.03$ & $-0.64$ & $-3.07$ & $-2.27$ & $-29.60$ & 2.09 & $-4.03$ & 3.09 & 6.05 & $-7.06$ & 49.99 &  \\ 
		\hline
	\end{tabular} \caption{Panel B PC OLS Regressions: Full specification regressions according to Equation \eqref{regression} including first PC of residuals, across loan term types} \label{tab:ols1pc2}
\end{table}

\end{landscape}

\begin{table}[ht]
	\centering
	\begin{tabular}{lrrrrr}
		\hline
		& CanCor & CanCor$^2$ & Eigenvalue & Percentage & Cum. Perc \\ 
		\hline
		1 & 0.99433 & 0.98868 & 87.39400 & 64.84569 & 64.85\\
		2 & 0.98681 & 0.97380 & 37.17146 & 27.58094 & 92.43\\
		3 & 0.93412 & 0.87258 & 6.84843 & 5.08148 & 97.51\\
		4 & 0.74667 & 0.55752 & 1.25998 & 0.93490 & 98.44\\
		5 & 0.67833 & 0.46012 & 0.85228 & 0.63239 & 99.08\\
		6 & 0.63521 & 0.40349 & 0.67643 & 0.50191 & 99.58\\
		7 & 0.51282 & 0.26298 & 0.35682 & 0.26476 & 99.84\\
		8 & 0.36017 & 0.12972 & 0.14905 & 0.11060 & 99.95\\
		9 & 0.23485 & 0.05515 & 0.05837 & 0.04331 & 100.00\\
		10 & 0.07318 & 0.00535 & 0.00538 & 0.00399 & 100.00\\
		\hline
	\end{tabular} \caption{Canonical Correlation Coefficients and Eigenvalues} \label{tab:ccaeigen}
\end{table}

\begin{table}[ht]
	\centering
	\begin{tabular}{lrrrrrr}
		\hline
		& CanCor & LR test stat & approx F & numDF & denDF & Pr($>$ F) \\ 
		\hline
		1 & 0.99 & 0.00 & 11.36 & 120.00 & 332.93 & 0.0000 \\ 
		2 & 0.99 & 0.00 & 6.77 & 99.00 & 307.63 & 0.0000 \\ 
		3 & 0.93 & 0.01 & 3.65 & 80.00 & 281.29 & 0.0000 \\ 
		4 & 0.75 & 0.09 & 2.20 & 63.00 & 253.92 & 0.0000 \\ 
		5 & 0.68 & 0.19 & 1.86 & 48.00 & 225.48 & 0.0015 \\ 
		6 & 0.64 & 0.36 & 1.54 & 35.00 & 195.93 & 0.0355 \\ 
		7 & 0.51 & 0.60 & 1.07 & 24.00 & 165.17 & 0.3778 \\ 
		8 & 0.36 & 0.82 & 0.67 & 15.00 & 132.91 & 0.8105 \\ 
		9 & 0.23 & 0.94 & 0.39 & 8.00 & 98.00 & 0.9256 \\ 
		10 & 0.07 & 0.99 & 0.09 & 3.00 & 50.00 & 0.9654 \\ 
		\hline
	\end{tabular}
	\caption{Wilk's Lambda Tests for Significance of Canonical Variates} \label{tab:lambdatest}
\end{table}

\begin{table}[ht]
	\centering
	\begin{tabular}{rrrrrrrrrrr}
		\hline
		$Y$ Variate & 1 & 2 & 3 & 4 & 5 & 6 & 7 & 8 & 9 & 10 \\ 
		\hline
		Redundancy & 0.37 & 0.16 & 0.30 & 0.01 & 0.01 & 0.00 & 0.00 & 0.00 & 0.00 & 0.00 \\ 
		\hline
	\end{tabular}\caption{Redundancy Indices for P2P Credit Spread Canonical Variates} \label{tab:redunindices}
\end{table}

\begin{table}[ht]
	\centering
	\begin{tabular}{rrrr}
		\hline
		& $Y$-Can 1 & $Y$-Can 2 & $Y$-Can 3 \\ 
		\hline
		CPI & 0.72 & 0.65 & 0.15 \\ 
		UNRATE & $-0.63$ & $-0.73$ & 0.15 \\ 
		Debt & $-0.84$ & $-0.44$ & $-0.21$ \\ 
		GDP & $-0.11$ & 0.56 & $-0.17$ \\ 
		SPX & 0.50 & 0.84 & $-0.11$ \\ 
		$\Delta$ 10Y-rf & $-0.01$ & 0.05 & $-0.00$ \\ 
		rf-slope & $-0.89$ & 0.35 & 0.22 \\ 
		SMB & $-0.16$ & $-0.06$ & $-0.21$ \\ 
		HML & 0.09 & $-0.14$ & 0.19 \\ 
		$\Delta$VIX & $-0.00$ & 0.11 & 0.01 \\ 
		\hline
	\end{tabular}\caption{Pairwise Correlations of CCA Factors and Macroeconomic Proxies} \label{tab:pairwisecorr}
\end{table}

\begin{table}[ht]
	\centering
	\begin{tabular}{rrrrrr}
		\hline
		& (Intercept) & Factor1 & Factor2 & Factor3 & Adj.\ $R^2$ \\ 
		\hline
		A & 6.71 & $\mathbf{0.37}$ & $\mathbf{-0.29}$ & $\mathbf{0.46}$ & 0.76 \\ 
		& 207.99 & 11.39 & $-8.79$ & 14.24 &  \\ 
		B & 10.16 & $\mathbf{0.40}$ & $\mathbf{-0.66}$ & $\mathbf{0.72}$ & 0.86 \\ 
		& 272.01 & 10.54 & $-17.49$ & 19.14 &  \\ 
		C & 13.23 & $\mathbf{0.57}$ & $\mathbf{-0.62}$ & $\mathbf{0.76}$ & 0.82 \\ 
		& 280.53 & 11.93 & $-12.95$ & 15.90 &  \\ 
		D & 16.15 & $\mathbf{0.97}$ & $\mathbf{-0.47}$ & $\mathbf{0.82}$ & 0.84 \\ 
		& 304.90 & 18.13 & $-8.85$ & 15.27 &  \\ 
		E & 18.65 & $\mathbf{1.26}$ & $\mathbf{-0.14}$ & $\mathbf{1.12}$ & 0.86 \\ 
		& 308.11 & 20.67 & $-2.28$ & 18.33 &  \\ 
		F & 21.47 & $\mathbf{1.73}$ & $\mathbf{0.70}$ & $\mathbf{0.77}$ & 0.93 \\ 
		& 441.27 & 35.23 & 14.23 & 15.65 &  \\ 
		\hline
	\end{tabular} \caption{Panel A CCA Factor Regressions: specification according to Equation \eqref{ccaregression}, across loan grade types} \label{tab:ccafactor}
\end{table}
\begin{table}[ht]
	\centering
	\begin{tabular}{rrrrrr}
		\hline
		& (Intercept) & Factor1 & Factor2 & Factor3 & Adj.\ $R^2$ \\
		\hline
		36-m & 14.55 & \textbf{0.77} & $-0.20$ & \textbf{0.77} & 0.04 \\ 
		& 54.51 & 2.86 & $-0.73$ & 2.86 &  \\ 
		60-m & 14.60 &  \textbf{0.82} & $-0.20$ & \textbf{0.80} & 0.04 \\ 
		& 55.57 & 3.08 & $-0.74$ & 3.02 &  \\ 
		\hline
	\end{tabular}\caption{Panel B CCA Factor Regressions: specification according to Equation \eqref{ccaregression}, across loan term types} \label{tab:ccafactor2}
\end{table}

\begin{table}[ht]
	\centering
	\begin{tabular}{rrrrrrr}
		\hline
		& (Intercept) & Factor1 & Factor2 & Factor3 & PC1 & Adj.\ $R^2$ \\
		\hline
		A & 6.71 & 0.37 & $-0.29$ & 0.46 & $-0.08$ & 0.81 \\ 
		& 233.05 & 12.76 & $-9.85$ & 15.96 & $-5.67$ &  \\ 
		B & 10.16 & 0.40 & $-0.66$ & 0.72 & $-0.16$ & 0.95 \\ 
		& 462.54 & 17.93 & $-29.74$ & 32.54 & $-15.22$ &  \\ 
		C & 13.23 & 0.57 & $-0.62$ & 0.76 & $-0.23$ & 0.97 \\ 
		& 684.63 & 29.11 & $-31.59$ & 38.80 & $-24.61$ &  \\ 
		D & 16.14 & 0.97 & $-0.47$ & 0.81 & $-0.26$ & 0.97 \\ 
		& 759.25 & 45.15 & $-22.05$ & 38.03 & $-25.21$ &  \\ 
		E & 18.65 & 1.26 & $-0.14$ & 1.12 & $-0.31$ & 0.99 \\ 
		& 1112.31 & 74.61 & $-8.22$ & 66.18 & $-38.33$ &  \\ 
		F & 21.47 & 1.73 & 0.70 & 0.77 & $-0.24$ & 0.99 \\ 
		& 1489.26 & 118.89 & 48.01 & 52.81 & $-35.62$ &  \\ 
		\hline
	\end{tabular} \caption{Panel A CCA-PC Factor Regressions: specification according to Equation \eqref{ccaregression} including first PC of residuals, across loan grade types} \label{tab:ccaregression}
\end{table}

\begin{table}[ht]
	\centering
	\begin{tabular}{rrrrrrr}
		\hline
		& (Intercept) & Factor1 & Factor2 & Factor3 & PC1 & Adj.\ $R^2$ \\
		\hline
		1 & 14.55 & 0.77 & $-0.20$ & 0.77 & 3.65 & 0.58 \\ 
		2 & 82.75 & 4.35 & $-1.10$ & 4.35 & 22.11 &  \\ 
		3 & 14.60 & 0.82 & $-0.20$ & 0.80 & 3.59 & 0.58 \\ 
		4 & 84.37 & 4.68 & $-1.12$ & 4.58 & 22.11 &  \\ 
		\hline
	\end{tabular} \caption{Panel B CCA-PC Factor Regressions: specification according to Equation \eqref{ccaregression} including first PC of residuals, across loan term types} \label{tab:ccaregression2}
\end{table}

\clearpage

\bibliographystyle{apacite}
\bibliography{bibcca}

\begin{thebibliography}{}

\bibitem [\protect \citeauthoryear {%
Ahn%
, Dieckmann%
\BCBL {}\ \BBA {} Perez%
}{%
Ahn%
\ \protect \BOthers {.}}{%
{\protect \APACyear {2012}}%
}]{%
ahn}
\APACinsertmetastar {%
ahn}%
\begin{APACrefauthors}%
Ahn, S\BPBI C.%
, Dieckmann, S.%
\BCBL {}\ \BBA {} Perez, M\BPBI F.%
\end{APACrefauthors}%
\unskip\
\newblock
\APACrefYearMonthDay{2012}{}{}.
\newblock
\APACrefbtitle {Exploring common factors in the term structure of credit
  spreads: the use of canonical correlations.} {Exploring common factors in the
  term structure of credit spreads: the use of canonical correlations.}
\newblock
\APACrefnote{Available at: \url{https://ssrn.com/abstract=984245}}
\PrintBackRefs{\CurrentBib}

\bibitem [\protect \citeauthoryear {%
Ang%
\ \BBA {} Piazzesi%
}{%
Ang%
\ \BBA {} Piazzesi%
}{%
{\protect \APACyear {2003}}%
}]{%
ang}
\APACinsertmetastar {%
ang}%
\begin{APACrefauthors}%
Ang, A.%
\BCBT {}\ \BBA {} Piazzesi, M.%
\end{APACrefauthors}%
\unskip\
\newblock
\APACrefYearMonthDay{2003}{}{}.
\newblock
{\BBOQ}\APACrefatitle {A no-arbitrage vector autoregression of term structure
  dynamics with macroeconomic and latent variables} {A no-arbitrage vector
  autoregression of term structure dynamics with macroeconomic and latent
  variables}.{\BBCQ}
\newblock
\APACjournalVolNumPages{Journal of Monetary economics}{50}{4}{745--787}.
\PrintBackRefs{\CurrentBib}

\bibitem [\protect \citeauthoryear {%
Bergbrant%
\ \BBA {} Kelly%
}{%
Bergbrant%
\ \BBA {} Kelly%
}{%
{\protect \APACyear {2016}}%
}]{%
berg}
\APACinsertmetastar {%
berg}%
\begin{APACrefauthors}%
Bergbrant, M\BPBI C.%
\BCBT {}\ \BBA {} Kelly, P\BPBI J.%
\end{APACrefauthors}%
\unskip\
\newblock
\APACrefYearMonthDay{2016}{}{}.
\newblock
{\BBOQ}\APACrefatitle {Macroeconomic expectations and the size, value, and
  momentum factors} {Macroeconomic expectations and the size, value, and
  momentum factors}.{\BBCQ}
\newblock
\APACjournalVolNumPages{Financial Management}{45}{4}{809--844}.
\PrintBackRefs{\CurrentBib}

\bibitem [\protect \citeauthoryear {%
Dietrich%
\ \BBA {} Wernli%
}{%
Dietrich%
\ \BBA {} Wernli%
}{%
{\protect \APACyear {2016}}%
}]{%
dietrich}
\APACinsertmetastar {%
dietrich}%
\begin{APACrefauthors}%
Dietrich, A.%
\BCBT {}\ \BBA {} Wernli, R.%
\end{APACrefauthors}%
\unskip\
\newblock
\APACrefYearMonthDay{2016}{}{}.
\newblock
\APACrefbtitle {What Drives the Interest Rates in the P2P Consumer Lending
  Market? Empirical Evidence from Switzerland.} {What drives the interest rates
  in the p2p consumer lending market? empirical evidence from switzerland.}
\newblock
\APACrefnote{Available at: \url{https://ssrn.com/abstract=2767455}}
\PrintBackRefs{\CurrentBib}

\bibitem [\protect \citeauthoryear {%
Duffee%
}{%
Duffee%
}{%
{\protect \APACyear {1998}}%
}]{%
duffee}
\APACinsertmetastar {%
duffee}%
\begin{APACrefauthors}%
Duffee, G\BPBI R.%
\end{APACrefauthors}%
\unskip\
\newblock
\APACrefYearMonthDay{1998}{}{}.
\newblock
{\BBOQ}\APACrefatitle {The relation between treasury yields and corporate bond
  yield spreads} {The relation between treasury yields and corporate bond yield
  spreads}.{\BBCQ}
\newblock
\APACjournalVolNumPages{The Journal of Finance}{53}{6}{2225--2241}.
\PrintBackRefs{\CurrentBib}

\bibitem [\protect \citeauthoryear {%
Fama%
\ \BBA {} French%
}{%
Fama%
\ \BBA {} French%
}{%
{\protect \APACyear {1993}}%
}]{%
fama}
\APACinsertmetastar {%
fama}%
\begin{APACrefauthors}%
Fama, E\BPBI F.%
\BCBT {}\ \BBA {} French, K\BPBI R.%
\end{APACrefauthors}%
\unskip\
\newblock
\APACrefYearMonthDay{1993}{}{}.
\newblock
{\BBOQ}\APACrefatitle {Common risk factors in the returns on stocks and bonds}
  {Common risk factors in the returns on stocks and bonds}.{\BBCQ}
\newblock
\APACjournalVolNumPages{Journal of financial economics}{33}{1}{3--56}.
\PrintBackRefs{\CurrentBib}

\bibitem [\protect \citeauthoryear {%
Gourieroux%
, Monfort%
\BCBL {}\ \BBA {} Renault%
}{%
Gourieroux%
\ \protect \BOthers {.}}{%
{\protect \APACyear {1993}}%
}]{%
gourieroux}
\APACinsertmetastar {%
gourieroux}%
\begin{APACrefauthors}%
Gourieroux, C.%
, Monfort, A.%
\BCBL {}\ \BBA {} Renault, E.%
\end{APACrefauthors}%
\unskip\
\newblock
\APACrefYearMonthDay{1993}{}{}.
\newblock
{\BBOQ}\APACrefatitle {Indirect inference} {Indirect inference}.{\BBCQ}
\newblock
\APACjournalVolNumPages{Journal of applied econometrics}{8}{S1}{}.
\PrintBackRefs{\CurrentBib}

\bibitem [\protect \citeauthoryear {%
Hair%
\ \protect \BOthers {.}}{%
Hair%
\ \protect \BOthers {.}}{%
{\protect \APACyear {1998}}%
}]{%
hair}
\APACinsertmetastar {%
hair}%
\begin{APACrefauthors}%
Hair, J\BPBI F.%
, Black, W\BPBI C.%
, Babin, B\BPBI J.%
, Anderson, R\BPBI E.%
, Tatham, R\BPBI L.%
\BCBL {}\ \BOthersPeriod {.}\end{APACrefauthors}%
\unskip\
\newblock
\APACrefYear{1998}.
\newblock
\APACrefbtitle {Multivariate data analysis} {Multivariate data analysis}\
  (\BVOL~5)\ (\BNUM~3).
\newblock
\APACaddressPublisher{}{Prentice hall Upper Saddle River, NJ}.
\PrintBackRefs{\CurrentBib}

\bibitem [\protect \citeauthoryear {%
Herzenstein%
, Andrews%
, Dholakia%
\BCBL {}\ \BBA {} Lyandres%
}{%
Herzenstein%
\ \protect \BOthers {.}}{%
{\protect \APACyear {2008}}%
}]{%
herzenstein}
\APACinsertmetastar {%
herzenstein}%
\begin{APACrefauthors}%
Herzenstein, M.%
, Andrews, R\BPBI L.%
, Dholakia, U\BPBI M.%
\BCBL {}\ \BBA {} Lyandres, E.%
\end{APACrefauthors}%
\unskip\
\newblock
\APACrefYearMonthDay{2008}{}{}.
\newblock
{\BBOQ}\APACrefatitle {The democratization of personal consumer loans?
  Determinants of success in online peer-to-peer lending communities} {The
  democratization of personal consumer loans? determinants of success in online
  peer-to-peer lending communities}.{\BBCQ}
\newblock
\APACjournalVolNumPages{Boston University School of Management Research
  Paper}{14}{6}{}.
\PrintBackRefs{\CurrentBib}

\bibitem [\protect \citeauthoryear {%
Klafft%
}{%
Klafft%
}{%
{\protect \APACyear {2008}}%
}]{%
klafft}
\APACinsertmetastar {%
klafft}%
\begin{APACrefauthors}%
Klafft, M.%
\end{APACrefauthors}%
\unskip\
\newblock
\APACrefYearMonthDay{2008}{}{}.
\newblock
\APACrefbtitle {Peer to peer lending: auctioning microcredits over the
  internet.} {Peer to peer lending: auctioning microcredits over the internet.}
\newblock
\APACrefnote{Available at: \url{https://ssrn.com/abstract=1352383}}
\PrintBackRefs{\CurrentBib}

\bibitem [\protect \citeauthoryear {%
Litterman%
\ \BBA {} Scheinkman%
}{%
Litterman%
\ \BBA {} Scheinkman%
}{%
{\protect \APACyear {1991}}%
}]{%
litterman}
\APACinsertmetastar {%
litterman}%
\begin{APACrefauthors}%
Litterman, R\BPBI B.%
\BCBT {}\ \BBA {} Scheinkman, J.%
\end{APACrefauthors}%
\unskip\
\newblock
\APACrefYearMonthDay{1991}{}{}.
\newblock
{\BBOQ}\APACrefatitle {Common factors affecting bond returns} {Common factors
  affecting bond returns}.{\BBCQ}
\newblock
\APACjournalVolNumPages{The Journal of Fixed Income}{1}{1}{54--61}.
\PrintBackRefs{\CurrentBib}

\bibitem [\protect \citeauthoryear {%
Okun%
}{%
Okun%
}{%
{\protect \APACyear {1963}}%
}]{%
okun}
\APACinsertmetastar {%
okun}%
\begin{APACrefauthors}%
Okun, A\BPBI M.%
\end{APACrefauthors}%
\unskip\
\newblock
\APACrefYear{1963}.
\newblock
\APACrefbtitle {Potential GNP: its measurement and significance} {Potential
  gnp: its measurement and significance}.
\newblock
\APACaddressPublisher{}{Yale University, Cowles Foundation for Research in
  Economics}.
\PrintBackRefs{\CurrentBib}

\bibitem [\protect \citeauthoryear {%
Tu%
, Yao%
\BCBL {}\ \BBA {} Zhang%
}{%
Tu%
\ \protect \BOthers {.}}{%
{\protect \APACyear {2015}}%
}]{%
tu}
\APACinsertmetastar {%
tu}%
\begin{APACrefauthors}%
Tu, Y.%
, Yao, Q.%
\BCBL {}\ \BBA {} Zhang, R.%
\end{APACrefauthors}%
\unskip\
\newblock
\APACrefYearMonthDay{2015}{}{}.
\newblock
\APACrefbtitle {Error-Correction Factor Models.} {Error-correction factor
  models.}
\newblock
\APACrefnote{Available at:
  \url{http://stats.lse.ac.uk/q.yao/qyao.links/paper/tyz2015.pdf}}
\PrintBackRefs{\CurrentBib}

\bibitem [\protect \citeauthoryear {%
Vassalou%
}{%
Vassalou%
}{%
{\protect \APACyear {2000}}%
}]{%
vas}
\APACinsertmetastar {%
vas}%
\begin{APACrefauthors}%
Vassalou, M.%
\end{APACrefauthors}%
\unskip\
\newblock
\APACrefYear{2000}.
\newblock
\APACrefbtitle {The {F}ama--{F}rench factors as proxies for fundamental
  economic risks} {The {F}ama--{F}rench factors as proxies for fundamental
  economic risks}\ (\BNUM~181).
\newblock
\APACaddressPublisher{}{Center on Japanese Economy and Business, Columbia
  Business School}.
\PrintBackRefs{\CurrentBib}

\bibitem [\protect \citeauthoryear {%
Zhang%
, Jia%
, Diao%
, Hai%
\BCBL {}\ \BBA {} Li%
}{%
Zhang%
\ \protect \BOthers {.}}{%
{\protect \APACyear {2016}}%
}]{%
zhang}
\APACinsertmetastar {%
zhang}%
\begin{APACrefauthors}%
Zhang, Y.%
, Jia, H.%
, Diao, Y.%
, Hai, M.%
\BCBL {}\ \BBA {} Li, H.%
\end{APACrefauthors}%
\unskip\
\newblock
\APACrefYearMonthDay{2016}{}{}.
\newblock
{\BBOQ}\APACrefatitle {Research on Credit Scoring by Fusing Social Media
  Information in Online Peer-to-Peer Lending} {Research on credit scoring by
  fusing social media information in online peer-to-peer lending}.{\BBCQ}
\newblock
\APACjournalVolNumPages{Procedia Computer Science}{91}{}{168--174}.
\PrintBackRefs{\CurrentBib}

\end{thebibliography}

\end{document}